\documentclass[aps,twocolumn,superscriptaddress,showpacs]{revtex4-1}
\usepackage{graphicx}
\usepackage{natbib}
\usepackage{color}
\usepackage{amsmath}
\usepackage{amssymb}
\usepackage{array}
\usepackage{multirow}
\usepackage{chemformula}
\usepackage{subcaption}

\begin{document}
\pdfoutput=1
\title{From magnetic order to quantum disorder: a $\mu$SR study of the Zn-barlowite series of $S={\frac{1}{2}}$ kagom{\'e} antiferromagnets, Zn$_{x}$Cu$_{4-x}$(OH)$_{6}$FBr}
\author{K.~Tustain}
\affiliation{Department of Chemistry and Materials Innovation Factory, University of Liverpool, 51 Oxford Street, Liverpool, L7 3NY, UK}
\author{B.~Ward-O'Brien}
\affiliation{Department of Chemistry and Materials Innovation Factory, University of Liverpool, 51 Oxford Street, Liverpool, L7 3NY, UK}
\author{F.~Bert}
\affiliation{Universit{\'e} Paris-Saclay, CNRS, Laboratoire de Physique des Solides, 91405, Orsay, France}
\author{T.~-H.~Han}
\affiliation{James Frank Institute and Department of Physics, University of Chicago, Chicago, Illinois 60637, USA}
\affiliation{Materials Science Division, Argonne National Laboratory, Argonne, Illinois 60439, USA}
\author{H.~Luetkens}
\affiliation{Laboratory for Muon Spin Spectroscopy, Paul Scherrer Institute, 5232 Villigen PSI, Switzerland}
\author{T.~Lancaster}
\affiliation{Department of Physics, Durham University, South Road, Durham, DH1 3LE, UK}
\author{B.~M.~Huddart}
\affiliation{Department of Physics, Durham University, South Road, Durham, DH1 3LE, UK}
\author{P.~J.~Baker}
\affiliation{ISIS Neutron and Muon Source, Rutherford Appleton Laboratory, Chilton, Didcot, Oxford, OX11 0QX, UK}
\author{L.~Clark}
\email{l.m.clark@bham.ac.uk}
\affiliation{Department of Chemistry and Materials Innovation Factory, University of Liverpool, 51 Oxford Street, Liverpool, L7 3NY, UK}
\affiliation{School of Chemistry, University of Birmingham, Edgbaston, Birmingham, B15 2TT, UK}



\date{\today}

\begin{abstract}
\noindent We report a comprehensive muon spectroscopy study of the Zn-barlowite series of $S={\frac{1}{2}}$ kagom{\'e} antiferromagnets, Zn$_x$Cu$_{4-x}$(OH)$_{6}$FBr, for $x=0.00$ to $0.99(1)$. By combining muon spin relaxation and rotation measurements with state-of-the-art density-functional theory muon-site calculations, we observe the formation of both $\mu$--F and $\mu$--OH complexes in Zn-barlowite. From these stopping sites, implanted muon spins reveal the suppression of long-range magnetic order into a possible quantum spin liquid state upon increasing concentration of Zn-substitution. In the parent compound ($x=0$), static long-range magnetic order below $T_{\mathsf{N}}=15$~K manifests itself in the form of spontaneous oscillations in the time-dependent muon asymmetry signal consistent with the dipolar fields expected from the calculated muon stopping sites and the previously determined magnetic structure of barlowite. Meanwhile, in the $x=1.0$ end-member of the series---in which antiferromagnetic kagom{\'e} layers of Cu$^{2+}$ $S={\frac{1}{2}}$ moments are decoupled by diamagnetic Zn$^{2+}$ ions---we observe that dynamic magnetic moment fluctuations persist down to at least 50 mK, indicative of a quantum disordered ground state. We demonstrate that this crossover from a static to dynamic magnetic ground state occurs for compositions of Zn-barlowite with $x>0.5$, which bears resemblance to dynamical behaviour of the widely studied Zn-paratacamite series that contains the quantum spin liquid candidate herbertsmithite. 
\end{abstract}
\maketitle

\begin{figure*}[!hbt]
\centering
\includegraphics[width=\linewidth]{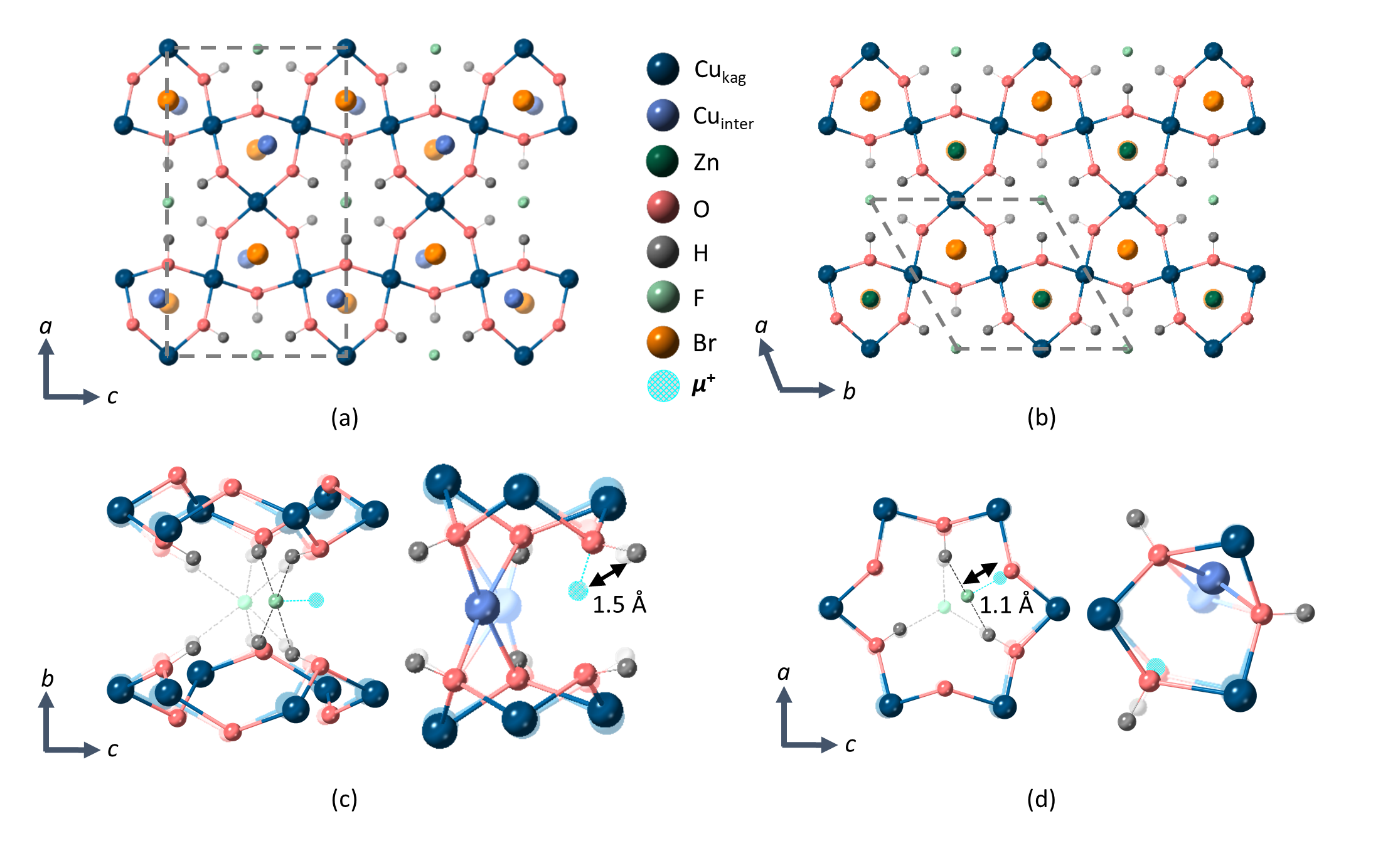}
\caption{\textbf{Low-temperature structures of barlowite and Zn-barlowite.} (a)~In barlowite, \ch{Cu4(OH)6FBr}, distorted kagom{\'e} layers---formed from two distinct Cu$^{2+}$ ions sites (Cu$_{\mathsf{kag}}$) and connected via hydroxide anions---are separated by a third, interlayer Cu$^{2+}$ ion site (Cu$_{\mathsf{inter}}$) shifted away from the centre of the triangular motifs in the kagom{\'e} layers above and below. Bromide and fluoride anions also lie between the kagom{\'e} planes. (b)~In Zn-barlowite, \ch{ZnCu3(OH)6FBr}, Zn$^{2+}$ ions separate undistorted kagom{\'e} layers formed from a single Cu$^{2+}$ ion site. The two classes of muon stopping sites calculated for barlowite, $\mu$-F (left) and $\mu$-OH (right), are shown here in the orthorhombic structure viewed along (c)~the $a$-axis, between the kagom{\'e} layers and (d)~along the $b$-axis perpendicular to the kagom{\'e} planes. Faded atoms indicate the unperturbed crystal lattice before muon implantation.}
\label{fig:1}
\end{figure*} 

\begin{figure}[!h]
\centering
\includegraphics[width=\linewidth]{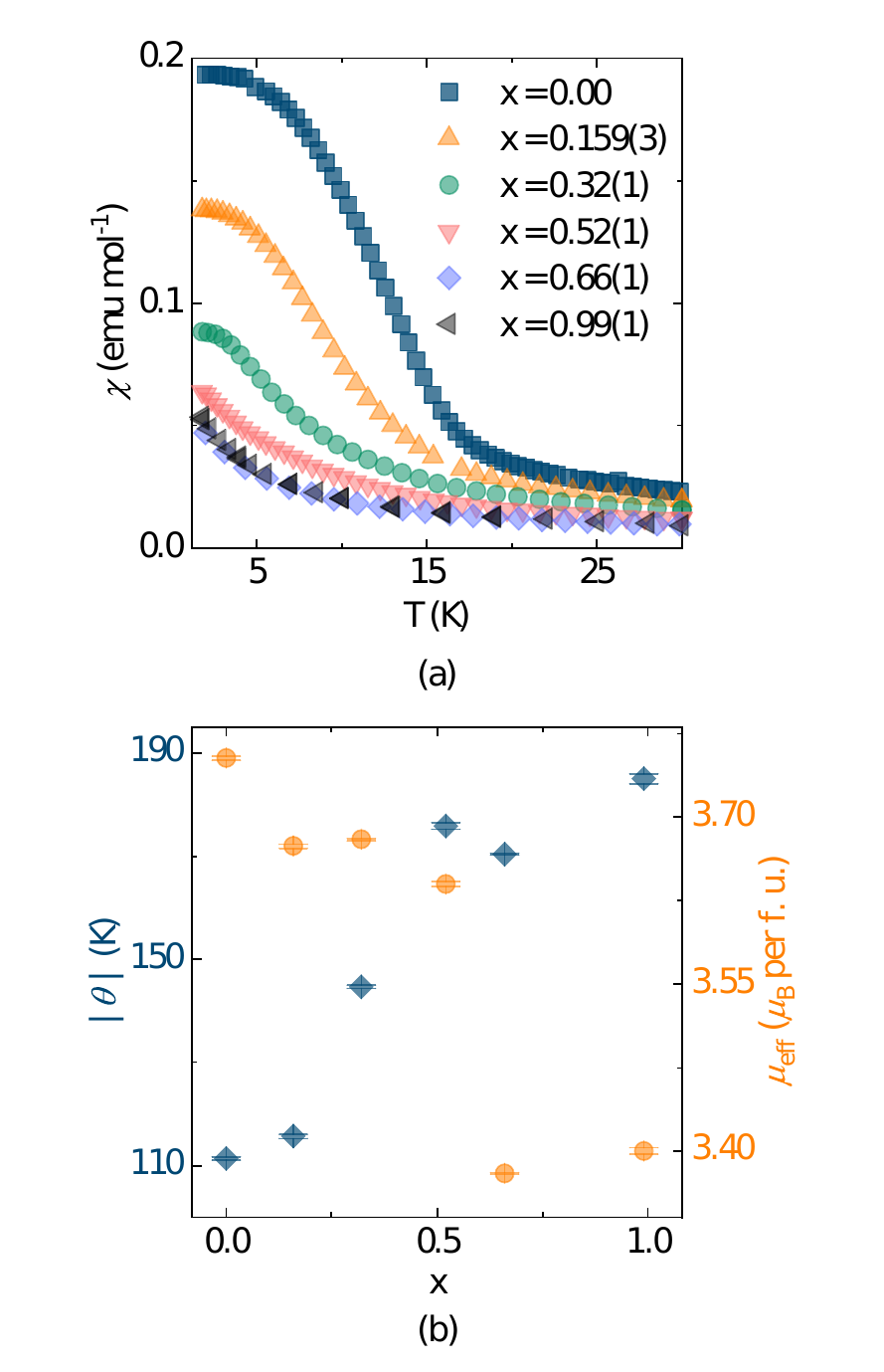}
\caption{\textbf{Magnetic susceptibility of Zn$_x$Cu$_{4-x}$(OH)$_{6}$FBr.} (a)~The temperature-dependent magnetic susceptibility measured in a 1 T field with zero-field cooling. (b)~The absolute values of the antiferromagnetic Weiss constants, $\theta$, and the effective magnetic moments, $\mu_{\mathsf{eff}}$, obtained from Curie-Weiss fitting the inverse magnetic susceptibility for all compositions, $x$, as shown in Fig. S1 of the Supplementary Information.}
\label{fig:2}
\end{figure}

\section{Introduction}
\noindent Quantum effects play a significant role in the low-temperature physics of magnetic systems in which antiferromagnetic $S=\frac{1}{2}$ moments decorate a two-dimensional kagom{\'e} array of corner-sharing triangles \cite{Greedan2001}. The competing interactions resulting from the geometric frustration in such systems combined with quantum fluctuations may give rise to elusive states of matter, such as quantum spin liquids (QSLs); a possibility that continues to capture the interest of the quantum materials research community \cite{Broholm2020}. Indeed, QSLs represent a unique state of matter that evades long-range magnetic order despite the sometimes considerable exchange interactions between its magnetic moments, which would typically drive a symmetry-breaking magnetic phase transition in a conventional magnet following Landau theory \cite{Knolle2019}. Instead, QSLs are quantum superpositional states formed from entangled pairs of magnetic moments, or spins, that can, in certain cases, be envisioned within a resonating valence bond model \cite{Anderson1973}. QSLs are of fundamental importance as their highly entangled nature should give rise to exotic physical phenomena, such as topological phases, fractionalised excitations and emergent gauge fields \cite{Savary2017}. However, the unambiguous realisation of a QSL remains an outstanding challenge.\\

In particular, while from a theoretical perspective it is generally accepted that the ground state of the $S=\frac{1}{2}$ kagom{\'e} antiferromagnet is a QSL \cite{Yan2011}, the true nature of this putative QSL state is still under debate, and new material realisations of it are needed to help resolve the ongoing disconnect between theory and experiment \cite{Han2012,Fu2015,Khuntia2020}. In this regard, synthetic analogues of Cu(II) hydroxyl halide minerals have been central to the experimental exploration into the physics of the $S=\frac{1}{2}$ kagom{\'e} antiferromagnet in recent years \cite{Fak2012,Boldrin2016,Hiroia}. To date, the most widely studied example of this class of materials is the Zn-paratacamite series, Zn$_{x}$Cu$_{4-x}$(OH)$_6$Cl$_2$, of which the $x=1$ end-member is the QSL candidate herbertsmithite, ZnCu$_{3}$(OH)$_6$Cl$_2$. The crystal structure of herbertsmithite contains kagom{\'e} layers of Cu$^{2+}$ $S=\frac{1}{2}$ moments separated by diamagnetic Zn$^{2+}$ ions \cite{Shores2005,Helton2007,DeVries2008}, which give rise to a dynamic, disordered magnetic ground state with magnetic moment fluctuations persisting to the lowest measurable temperatures despite  strong nearest-neighbour antiferromagnetic exchange (${J=200}$~K) \cite{Helton2007,Mendels2007}. Precisely characterising the instrinsic physics of the $S={\frac{1}{2}}$ kagom{\'e} layers of herbertsmithite, however, is extremely challenging due to the presence of Cu${^{2+}}$/Zn$^{2+}$ site occupancy disorder within its crystal structure \cite{DeVries2008,Freedman2010,Nilsen2013}. \\

A related Cu(II) hydroxyl halide that has garnered considerable attention in the recent literature is barlowite, Cu$_4$(OH)$_6$FBr, \cite{Elliott2014}, a naturally occurring mineral which has been synthetically produced by several groups \cite{Han2014,Feng,Pasco2018,Tustain2018,Smaha2018,Henderson2018}. A recent powder neutron diffraction (PND) study contributed to the ongoing debate around the nuclear and magnetic structures of this material \cite{Tustain2018}: like herbertsmithite, barlowite is formed from Cu$^{2+}$ $S={\frac{1}{2}}$ kagom{\'e} layers but with a second interlayer Cu$^{2+}$ site, which is disordered at room-temperature in the widely reported $P6_{3}/mmc$ model \cite{Han2014,Feng,Henderson2018,Smaha2018}. Below approximately 250 K, one reported class of sample undergoes a subtle structural distortion to an orthorhombic $Pnma$ phase \cite{Tustain2018} [see Fig. \ref{fig:1}(a)], which appears to relieve magnetic frustration and allows for the onset of magnetic order at $T_{\mathsf{N}}=15$ K \cite{Han2016}. Intriguingly, computational studies have suggested that barlowite may be tuned towards a QSL phase by substituting the interlayer Cu$^{2+}$ ions with diamagnetic Zn$^{2+}$ or Mg$^{2+}$ \cite{Liu2015,Guterding2016} thus reducing the magnetic coupling between the frustrated $S={\frac{1}{2}}$ kagom{\'e} layers. Furthermore, the calculated energy of formation for cation site occupancy defects is an order of magnitude greater in substituted barlowite than for herbertsmithite \cite{Liu2015}, suggesting Mg-barlowite or Zn-barlowite may be highly promising candidates in which to explore the instrisic properties of a Cu$^{2+}$-based $S={\frac{1}{2}}$ kagom{\'e} antiferromagnet. Indeed, the end-member of the Zn-barlowite series, nominally \ch{ZnCu3(OH)6FBr}, has been studied using inelastic neutron scattering and $^{19}$F NMR and seems to display evidence for a gapped Z$_{2}$ QSL \cite{Feng2017,Wei2019}, a theoretically predicted ground state for the $S=\frac{1}{2}$  Heisenberg model on an antiferromagnetic kagom{\'e} net \cite{Depenbrocka}.\\

Despite this promise, the evolution of the magnetic ground state and characteristic magnetic moment correlation dynamics across the Zn-barlowite series are yet to be explored in any great detail using local probe techniques, such as muon spin relaxation and rotation ($\mu$SR). In these techniques, spin-polarised, positively-charged muons are implanted into a sample of interest, in which they thermalise in regions of high electron density. From there, the spin polarisation of implanted muons will evolve over time, depending on the presence of local magnetic fields at the muon stopping site, which may originate from both nuclear and electronic moments within the sample. Following the time-dependence of the muon spin polarisation---via the asymmetry of the emmitted muon-decay positrons---thus provides a unique insight into the local magnetic properties of condensed matter systems \cite{Blundell1999}. As the muon spin is sensitive to very small magnetic fields ($\sim$ $10^{-5}$~T) and fluctuating field dynamics on a microsecond timescale, $\mu$SR has proved to be invaluable in revealing dynamical magnetic correlations in herbertsmithite and many other frustrated quantum magnets \cite{Mendels2007,Zheng2005,Kermarrec2011a,Fak2012,Barthelemy,Zorko2019,Clark2013}. However, limitations in the interpretation of ${\mu}$SR data often stem from the lack of knowledge of the possible muon stopping sites within complex crystal structures and the extent to which the implanted muons distort the surrounding lattice \cite{Foronda2015}. Crucially, the development of density-functional theory (DFT) to calculate muon stopping sites in crystalline solids is increasingly enabling the determination of these factors to advance the quantitative analysis of experimental ${\mu}$SR data \cite{Moller2013}. Here, we present such a study---combining comprehensive $\mu$SR measurements on the Zn-barlowite series with supporting DFT muon-site calculations---to reveal the onset of the possible QSL phase within this family of quantum materials. In this case, we take advantage of the formation of both $\mu-$F and $\mu-$OH complexes upon muon implantation in Zn-barlowite which---in combination with our DFT calculations---gives a strong constraint for the accurate determination of muon stopping sites and the local magnetic properties of this series.

\begin{figure*}[!t]
\centering
\includegraphics[width=\linewidth]{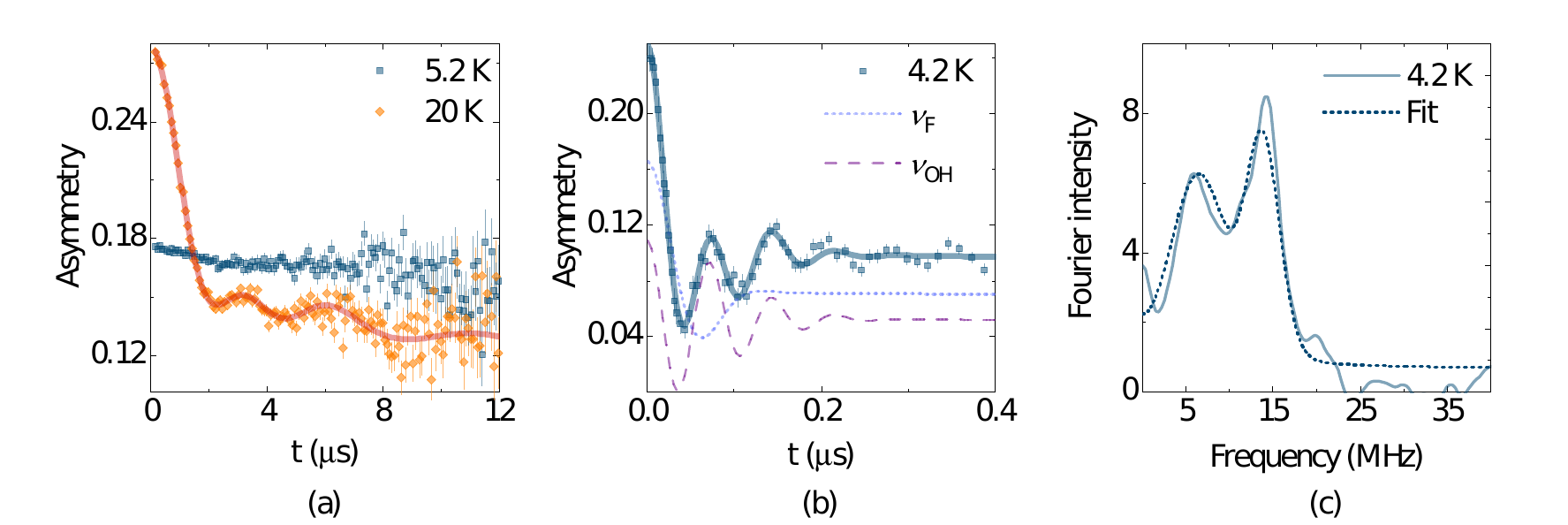}
\caption{\textbf{Time-dependent zero-field muon asymmetry for \ch{Cu4(OH)6FBr}}. (a)~Data measured at $5.2$~K and $20$~K on the MuSR spectrometer, with the solid line showing the fit of Equation $1$ to the high-temperature data. (b) ~Data collected at $4.2$~K on the GPS instrument, which can be modelled using Equation $4$, as shown by the solid line. The dashed lines indicate the two distinct oscillation frequencies contributing to the model. (c)~The Fourier transform of the time-dependent signal shown in part (b)~clearly reveals these two frequencies at $\nu_{\mathsf{F}}=6.3(2)$~MHz and $\nu_{\mathsf{OH}}=13.7(2)$~MHz, corresponding to local dipolar fields, $B^{\mathsf{loc}}_{\mathsf{F}}=46(2)$~mT and $B^{\mathsf{loc}}_{\mathsf{OH}}=101(2)$~mT.}
\label{fig:3}
\end{figure*}

\section{Results and Discussion}

\begin{table}
\caption{\label{label} \textbf{Muon stopping sites in \ch{Cu4(OH)6FBr}}. \textbf{Top:} Parameters obtained from fitting Equation 1 to zero-field, 20 K data collected on MuSR (Fig. \ref{fig:3}(a)). \textbf{Bottom:} The fraction of muons, $f$, stopping at each class of site, as determined from MuSR (Equation 1, \ref{fig:3}(a)) and GPS data (Equation 4, \ref{fig:3}(b)) as well as the comparison between experiment and calculation for the $\mu$--F or $\mu$--H bond lengths, $d$, and their relative local dipolar fields, $B^{\mathsf{loc}}$, below $T_{\mathsf{N}}$.}

\begin{ruledtabular}
\hspace{-0.75cm}
\begin{tabular}{@{}cccc}
$\mu^+$ Stopping Site & $\omega$ (Mrad s$^{-1}$) & $\Delta$ (mT)\\
\hline \vspace{-0.2cm} \\
F$^-$ & 1.23(2) & 0.43(4)\\
OH$^-$ & 0.72(2) & 0.18(4)
\label{table:1}
\end{tabular}
\vspace{0.5cm}
\begin{tabular}{@{}cccccc}
\multicolumn{2}{c}{$f$ (\%)}&\multicolumn{2}{c}{$d$ (\AA{})}&\multicolumn{2}{c}{$B^{\mathsf{loc}}$ (mT)}\\
MuSR & GPS & MuSR & DFT & GPS & Calculated \\
\hline \vspace{-0.2cm} \\
69(4) & 64(2) & 1.22(1) & 1.11 & 46(2) & 55\\
31(4) & 34(2) & 1.50(3) & 1.54 & 101(2) & 120
\label{table:1}
\end{tabular}

\end{ruledtabular}
\end{table}

\noindent We first consider the parent compound barlowite, \ch{Cu_4(OH)_6FBr}, which---in accordance with previous studies \cite{Han2014,Feng,Pasco2018,Smaha2018}---orders at $T_{\mathsf{N}}=15$~K as indicated by the sharp upturn in the magnetic susceptibility shown in Fig. \ref{fig:2}(a). Fig. \ref{fig:3}(a) shows the muon decay asymmetry measured in zero-field (ZF) on the MuSR spectrometer for such a sample. Above $T_{\mathsf{N}}$, the long-time oscillations observed in the data are characteristic of the local nuclear magnetic fields at the muon stopping sites within the sample. In particular, such oscillations are indicative of dipolar coupling between the muon and nuclear spins that leads to the formation of entangled state complexes. DFT muon-site calculations reveal two distinct classes of muon stopping site complexes in barlowite, which we show are consistent with the experimental ZF asymmetry signal shown in Fig. \ref{fig:3}(a). The first class of muon stopping site localises approximately $1.0$~\AA\ away from the oxygen atoms in the hydroxide groups that connect the Cu$^{2+}$ ions within the kagom{\'e} layers of barlowite. This forms a triangular $\mu$--OH complex, with the muon-proton distance found to be $1.54$~\AA\ in the lowest-energy sites, as shown in Fig. \ref{fig:1}(c). In the second class of muon stopping site, muons are found to localise near the fluoride anions in between the kagom{\'e} layers of barlowite, with a $\mu$--F separation of $1.1$~\AA, as depicted in Fig. \ref{fig:1}(d). Such $\mu$--F sites lie at substantially higher energies above the lowest energy $\mu$--OH sites in our calculations ($\approx 1$~eV). Although this suggests---on purely energetic grounds---that the formation of $\mu$--F complexes in barlowite is unlikely compared to $\mu$--OH, it is possible that during the stopping process, muons can be captured in these local potential minima.\\

Based on the results of these DFT calculations, we have devised a model that describes the ZF asymmetry of barlowite in the paramagnetic regime, incorporating the two distinct classes of muon stopping sites,

\begin{equation}
\begin{aligned}
A(t)=f_{\mathsf{F}}P_{\mathsf{F}}(t)\exp\Big(\frac{-\Delta_{\mathsf{F}}^{2}t^{2}}{2}\Big)\\+ f_{\mathsf{OH}}P_{\mathsf{OH}}(t)\exp\Big(\frac{-\Delta_{\mathsf{OH}}^{2} t^{2}}{2}\Big)\\ + A_{\mathsf{bg}}\exp({-\lambda_{\mathsf{bg}}t}).
\end{aligned}
\end{equation}

\noindent $A_{\mathsf{bg}}$ accounts for the background contribution of muons that stop outside the sample, which is weakly relaxing, and the expressions $P_{\mathsf{F}}(t)$ and $P_{\mathsf{OH}}(t)$ describe the dipolar interactions between the nuclear magnetic moments of fluoride and hydroxide groups and the muon spin at each class of stopping site within the sample, given by \cite{Lord2000},
 
\begin{equation}
\begin{aligned}
P_{\mathsf{F(OH)}}(t)=\frac{1}{6}+\frac{1}{3}\cos\Big(\frac{\omega_{\mathsf{F(OH)}}t}{2}\Big)\\+\frac{1}{6}\cos\Big(\omega_{\mathsf{F(OH)}}t\Big)+\frac{1}{3}\cos\Big(\frac{3\omega_{\mathsf{F(OH)}}t}{2}\Big).
\end{aligned}
\end{equation}

\noindent The oscillation frequencies $\omega_{\mathsf{F}}$ and $\omega_{\mathsf{OH}}$ are related to the respective $\mu$--F and $\mu$--H distances, $d_{\mathsf{F(H)}}$, through,

\begin{equation}
\omega_{\mathsf{F(OH)}}=\frac{\mu_{0}\hbar \gamma_\mu \gamma_{\mathsf{F(H)}}}{4\pi d_{\mathsf{F(H)}}^{3}},
\end{equation}

\noindent where $\mu_{0}$ is the permeability of free space and $\gamma_{\mu}= 2\pi \times 135.5$~MHz~T$^{-1}$, $\gamma_{\mathsf{F}} = 2\pi \times 40.1$~MHz~T$^{-1}$ and $\gamma_{\mathsf{H}} = 2\pi \times 42.6$~MHz~T$^{-1}$ are the gyromagnetic ratios of the muon spin and the fluorine and hydrogen nuclear spins, respectively. A fit of this model to ZF data collected for barlowite is shown in Fig. \ref{fig:3}(a) with the fitting parameters given in Table \ref{table:1}. Crucially, the experimentally determined $\mu$--F and $\mu$--H distances are in good agreement with those calculated from DFT. The Gaussian damping term in Equation 1 phenomenologically describes a possible distribution of nuclear fields surrounding the muon spin, with $\Delta_{\mathsf{F}}$ and $\Delta_{\mathsf{OH}}$ describing the field distributions at each class of stopping site. Indeed, DFT calculations also suggest that there are many possible stopping sites within the $\mu$--OH class, separated in total energy by less than $0.1$~eV. This reflects the possibility that muons localise near the oxygen atoms in a number of similar configurations. There is some variation in the $\mu$--H distances calculated for these similar stopping sites, with $1.63~$\AA\ found in a few of the higher-energy sites in this class. In all calculated cases of $\mu$--OH complex, however, the muon causes a distortion to the local crystal structure in the vicinity of its stopping site. In the lowest-energy sites of this class, the muon pushes the closest Cu$^{2+}$ ion away from the muon site by approximately $0.45$~\AA, as demonstrated in Fig. \ref{fig:1}(c) and (d). In contrast, little distortion of the local environment occurs close to the $\mu$--F sites.\\

Below $T_{\mathsf{N}}$, the internal electronic magnetic fields arising from the ordered Cu$^{2+}$ moments in barlowite rapidly depolarise the implanted muon spins. As a consequence, much of the initial asymmetry in our MuSR datasets is lost, as can be seen in Fig. \ref{fig:3}(a), owing to the limited time-resolution of the pulsed muon beam at the ISIS Neutron and Muon Source. We do, however, observe a `one-third tail' in the baseline asymmetry, a feature attributed to the presence of static magnetic moments in a polycrystalline sample, where on average, one-third of the implanted muon spins will be initially parallel to the local magnetic field, and whose polarisation will thus remain unaffected in the absence of dynamics \cite{Yaouanc2010}. To overcome this limitation, we turn to data collected for barlowite on the GPS instrument at PSI, shown in Fig. \ref{fig:3}(b), for which the continuous nature of the muon source provides sufficient time-resolution to capture the spontaneous oscillations apparent below $0.25~{\mu}$s in the ZF asymmetry signal. These oscillations are indicative of the muon spin precession around the static internal magnetic fields of electronic origin at a frequency given by $\nu_{i}=(\gamma_{\mu}/2\pi)B_{i}^{\mathsf{loc}}$, where $B_{i}^{\mathsf{loc}}$ is the local magnetic field at each muon stopping site, $i$, arising from the neighbouring ordered Cu$^{2+}$ moments. Consistent with the DFT calculations and the analysis of the ZF  asymmetry data collected above $T_{\mathsf{N}}$, the spontaneous oscillations in the ZF  asymmetry data collected below $T_{\mathsf{N}}$ in Fig. \ref{fig:3}(b) can also be modelled in terms of two distinct classes of muon stopping site with the following expression,

\begin{equation}
\begin{aligned}
A(t)=\sum_{i=1}^{2}a_{i}\Big[\frac{2}{3}\cos(2\pi\nu_{i}+\phi)\exp(-\sigma_{i}^2t^2)\\ +~\frac{1}{3}\exp(-\lambda t)\Big]+ A_{\mathsf{bg}}.
\end{aligned}
\end{equation}

\noindent The two dominant oscillating frequencies, $\nu_{1}=6.3(2)$~MHz and $\nu_{2}=13.7(2)$~MHz, extracted from this fit can also be clearly resolved in the Fourier transform of the GPS data collected at $4.2$~K, as shown in Fig. \ref{fig:3}(c). Comparing the fitted amplitudes of each component, $a_1=64(2)$\% and $a_2=34(2)$\%, with the stopping site fractions, $f_{\mathsf{F}}$ and  $f_{\mathsf{OH}}$, obtained from fitting the ZF data above $T_{\mathsf{N}}$ in Fig. \ref{fig:3}(a) (see Table \ref{table:1}), we assign the former to the $\mu$--F and the latter to the $\mu$--OH classes of muon stopping site. From this assignment, the data imply that the local magnetic fields at each class of muon stopping site arising from neighbouring ordered Cu$^{2+}$ moments are thus $B_{\mathsf{F}}^{\mathsf{loc}}~{\approx}~46$~mT and $B_{\mathsf{OH}}^{\mathsf{loc}}~\approx~101$~mT, respectively. It should be noted that the weak relaxation of the one-third tail in the asymmetry data should not be mistaken for residual spin dynamics in barlowite below $T_{\mathsf{N}}$. In the case of Fig. \ref{fig:3}(b), it is due to the small fraction of muons that stop in the aluminium foil used to contain the sample, whose spins thus experience the nuclear fields of $^{27}$Al.\\

As a final consistency check for our combination of DFT methods and muon experiment, we estimate the local electronic magnetic moment dipolar fields at each of the calculated muon stopping site classes based on the known magnetic structure for barlowite below $T_{\mathsf{N}}$ \cite{Tustain2018}. For the lowest-energy $\mu$--OH stopping sites, and assuming an undistorted crystal structure, these calculations suggest typical local fields around $120$~mT, along with a few around $75$~mT. If the muon-induced distortion is taken into account we obtain still smaller fields on the order of $50$~mT, owing to the movement of the closest Cu$^{2+}$ moment. The $B_{\mathsf{OH}}^{\mathsf{loc}}$ fields obtained are, therefore, highly sensitive to this distortion, but it is probable that this effect is over estimated in our computations owing to the relatively small size of the supercell used.  The average $B_{\mathsf{F}}^{\mathsf{loc}}$ for the $\mu$--F class of stopping sites is smaller, at  typically $55$~mT, and these values are insensitive to the inclusion of the distortions. Overall, the agreement of these calculations with the fields determined from our PSI data (summarised in Table {\ref{table:1}}) reinforces the likely formation of $\mu$--F stopping sites in barlowite despite their overall higher energy compared with $\mu$--OH.\\

\begin{figure}[!b]
\centering
\includegraphics[width=\linewidth]{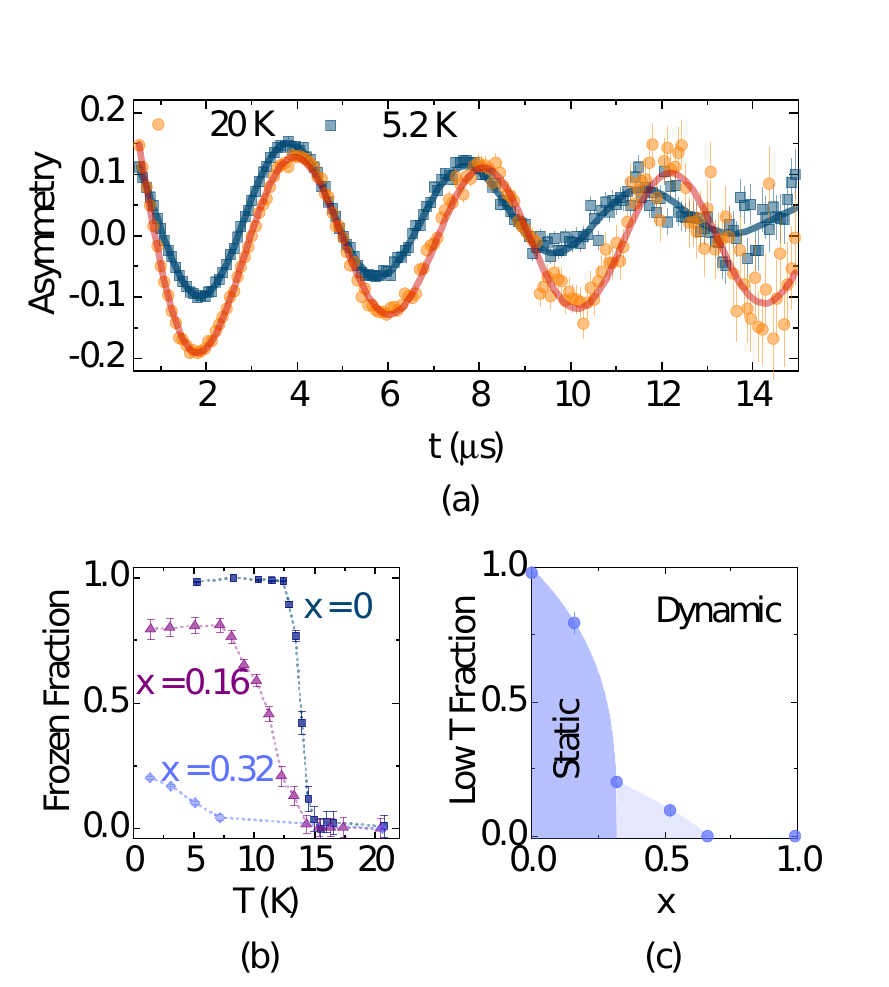}
\caption{\textbf{Time-dependent transverse-field muon asymmetry for \ch{Cu4(OH)6FBr}}. (a)~Asymmetry measured in a $2$~mT transverse-field at $5.2$ K and $20$ K. The solid lines are fits of Equation $5$ to the data. (b)~Frozen fraction of static Cu$^{2+}$ moments as a function of temperature determined from transverse-field data collected for a range of sample compositions, Zn$_x$Cu$_{4-x}$(OH)$_6$FBr. The dashed lines are guides to the eye. (c)~The low-temperature frozen fraction as a function of $x$, where the dark shaded region highlights the critical-like fall-off of static Cu$^{2+}$ moments upon the Zn$^{2+}$ substitution of interlayer sites. The lightly shaded region indicates a crossover regime between static and dynamic ground states.}
\label{fig:4}
\end{figure}

We next consider the effect of substituting \ch{Cu^{2+}} $S=\frac{1}{2}$ ions in barlowite for diamagnetic \ch{Zn^{2+}} in the series, Zn$_x$Cu$_{4-x}$(OH)$_6$FBr. Our analysis of powder neutron diffraction (PND) data collected for Zn-substituted samples, shown in Fig. S2 and Table S2 of the Supplementary Information (SI), yields two important observations. First, for highly-substituted samples ($x>0.5$) the crystal structure retains hexagonal $P6_3/mmc$ symmetry down to $1.5$ K---the base temperature of our diffraction experiments---in which the kagom{\'e} layers remain undistorted. Second, refinement of the cation site occupancies within this hexagonal structure indicate that the substituted Zn$^{2+}$ ions occupy the trigonal prismatic interplane site---in keeping with theory \cite{Guterding2016}---thus reducing the magnetic coupling between the kagom{\'e} layers of Cu$^{2+}$ moments. Correspondingly, we observe a progressive suppression of the magnetic ordering transition in the temperature-dependent magnetic susceptibility for Zn$_x$Cu$_{4-x}$(OH)$_6$FBr with increasing $x$, as shown in Fig. \ref{fig:2}(a). Fig. \ref{fig:2}(b) shows the fitting parameters obtained from the Curie-Weiss analysis of these data, revealing a reduction in the effective magnetic moment, $\mu_{\mathsf{eff}}$, per formula unit across the series as Cu$^{2+}$ is replaced with Zn$^{2+}$, and an increase in the absolute value of the Weiss constant, $\theta$, since the net magnetic exchange becomes increasingly negative as the positive ferromagnetic contribution from the interlayer coupling is reduced across the series.\\

In order to follow the evolution of the magnetic ground state in Zn-barlowite, we performed a set of transverse-field (TF) $\mu$SR measurements across the series, in which a magnetic field of $2$~mT is applied perpendicular to the initial muon spin polarisation. In this TF geometry for a sample in the paramagnetic regime, implanted muon spins precess around the applied field with a frequency, $\nu_{\mathsf{TF}}$. Below $T_{\mathsf{N}}$, however, the local internal magnetic fields within the magnetically ordered state can dominate the muon spin relaxation, resulting in a loss of asymmetry and a dephasing of the oscillating signal measured in the TF experiment. Both of these effects can be seen in the TF data collected for an $x=0.00$ sample above and below $T_{\mathsf{N}}$ shown in Fig. \ref{fig:4}(a), and can be modelled by,

\begin{equation}
\begin{aligned}
A(t)=[A_{\mathsf{p}}\exp(-\lambda_{\mathsf{TF}} t) +A_{\mathsf{bg}}\exp(-\sigma^{2}_{\mathsf{bg}} t^{2})]\\ \times \cos(2 \pi \nu_{\mathsf{TF}} t+ \phi)+B,
\end{aligned}
\end{equation}

\noindent where $A_{\mathsf{p}}$ provides a measure of the paramagnetic (dynamic) volume fraction of the sample, $A_{\mathsf{bg}}$ gives the background contribution of muons that stop in the silver sample holder and $B$ accounts for the fraction of muons with their spin polarisation aligned with the local magnetic fields in the ordered state of our polycrystalline samples. Therefore, in the $x=0.00$ sample below $T_{\mathsf{N}}$, $B$ tends to one-third of the asymmetry we observe in the paramagnetic state, in a similar manner to the one-third tail observed in our ZF data discussed above. By fitting this model to TF data collected over a range of temperatures, we may plot the temperature dependence of $A_{\mathsf{p}}$ or, in this case, the normalised inverse of this value (\textit{i.e.} $1-A_{\mathsf{p}}(T)/A_{\mathsf{p}}(20)$), which represents the evolution of the frozen volume fraction of static magnetic moments within a sample. For example, for the $x=0.00$ sample, a sudden drop in $A_{\mathsf{p}}$ at $T_{\mathsf{N}}=15$ K corresponds to a dramatic increase in the static volume fraction, shown in Fig. \ref{fig:4}(b), a clear signal of the magnetic ordering transition, in agreement with our magnetic susceptibility [Fig. \ref{fig:2}(a)] and PND measurements \cite{Tustain2018}.\\

The same analysis can be applied to samples in the Zn-barlowite series upon increasing Zn$^{2+}$ content, and Fig. \ref{fig:4}(b) shows the temperature dependence of the frozen volume fraction for Zn$_x$Cu$_{4-x}$(OH)$_6$FBr with $x=0.159(3)$ and $0.32(1)$. For $x=0.159(3)$  we observe an extended magnetic ordering regime, while for the $x=0.32(1)$ sample, the static moment fraction at base temperature is vastly reduced compared with the parent $x=0.00$ compound. Moreover, there is also evidence to support the coexistence of static and dynamic magnetic correlations in the $x=0.159(3)$ and $0.32(1)$ members of the Zn-barlowite series, as in both ZF and TF measurements of these samples we do not recover the one-third asymmetry expected for a fully ordered magnetic ground state in a polycrystalline sample, as shown in Fig. S3 of the Supplementary Information. As the Zn$^{2+}$ content of Zn$_x$Cu$_{4-x}$(OH)$_6$FBr is increased further to $x=0.52(1)$, dynamic magnetic correlations dominate the ZF muon asymmetry at low-temperature---likely indicating a static to dynamic crossover region in the magnetic phase diagram---and by $x=0.66(1)$, the magnetic moments of the Cu$^{2+}$ $S={\frac{1}{2}}$ ions remain dynamically fluctuating down to the lowest temperatures of our ${\mu}$SR experiment. This is most clearly evidenced by the low-temperature ZF data collected for the $x=0.66(1)$ sample shown in Fig. \ref{fig:5}(a), which can be modelled using the same expression developed for the $x=0.00$ compound in the paramagnetic regime (Equation 1 and Table S3 of the Supplementary Information). An important result from our DFT calculations in this regard is that the inclusion of Zn$^{2+}$ ions and the hexagonal space group symmetry of the Zn-substituted samples does not yield any notably different classes of low-energy muon stopping sites compared with the parent orthorhombic structure. Together, our observations from ZF and TF ${\mu}$SR measurements from across the Zn-barlowite series allow us to map out the low-temperature magnetic phase diagram shown in Fig. {\ref{fig:4}(c), which reveals a critical reduction in the static volume fraction from below $x=0.5$ as the system enters a possible QSL state.\\

\begin{figure}[!h]
\centering
\includegraphics[width=\linewidth]{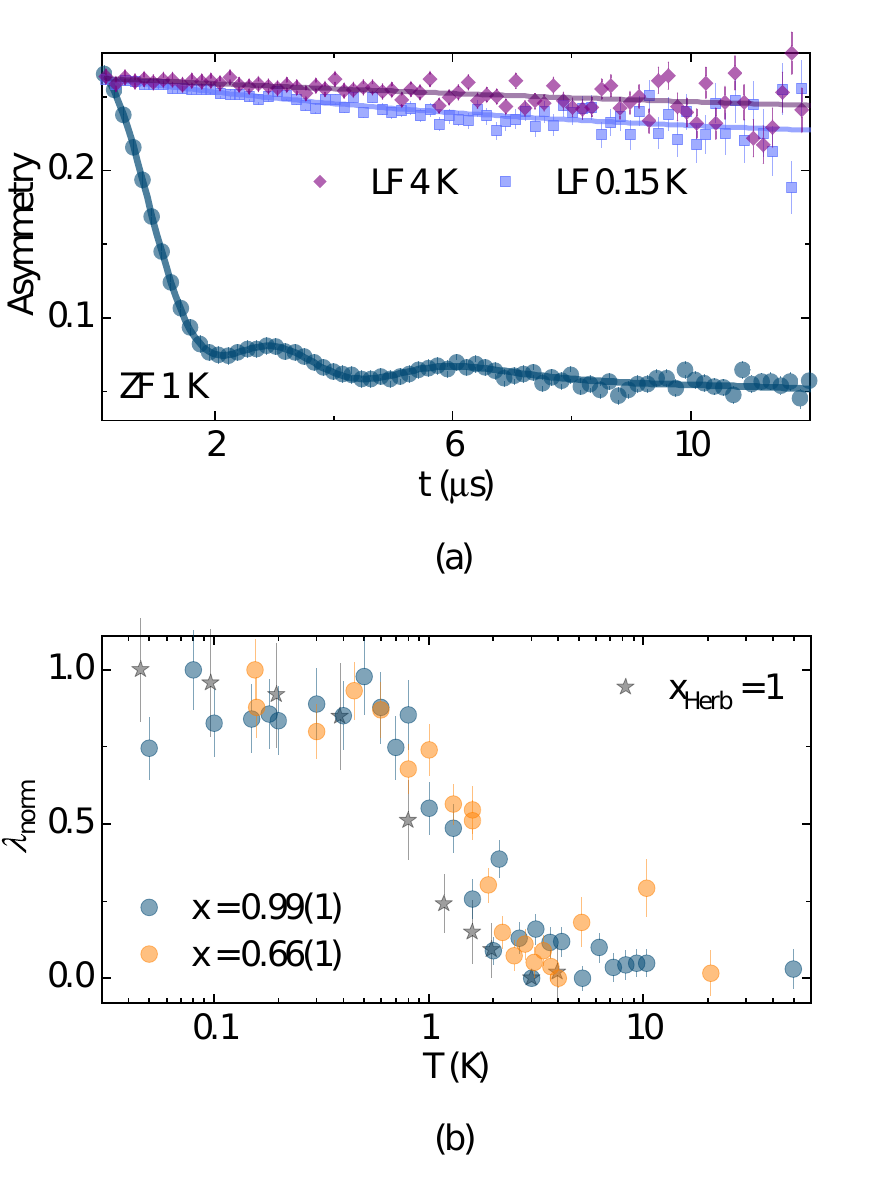}
\caption{\textbf{Persistent dynamics in Zn-barlowite.} (a)~Time-dependent muon asymmetry measured for Zn$_{0.66(1)}$Cu$_{3.34}$(OH)$_6$FBr in zero-field at $1$~K, where the solid line shows a fit using Equation 1, and in an applied longitudinal field of $10$~mT at $0.15$~K and $4$~K, where solid lines show fits of Equation 6 to the data. (b)~Temperature dependence of the normalised muon spin relaxation rate, $\lambda_{\mathsf{norm}}$, obtained from longitudinal-field data collected for $x=0.66(1)$ and $x=0.99(1)$ samples. As a comparison, data points for $x=1.0$ herbertsmithite, \ch{ZnCu3(OH)6Cl2}, obtained from  Ref.~\cite{Mendels2007}, are also shown.}
\label{fig:5}
\end{figure} 

Finally, we turn to explore the dynamical correlations within the quantum disordered ground state of 
Zn$_x$Cu$_{4-x}$(OH)$_6$FBr for $x\geq0.66$. In order to isolate and probe the spin dynamics of the electronic magnetic moments in these highly-substituted samples, we applied a longitudinal field (LF) of $10$~mT in our ${\mu}$SR experiments. In doing so, the implanted muon spins are decoupled from the the nuclear magnetic moments that dominate the ZF signal, meaning any remaining depolarisation of the muon spin arises from its interaction with the local electronic magnetic fields (see Fig. S5 in the Supplementary Information). Fig. \ref{fig:5}(a) shows representative LF data for the $x=0.66(1)$ sample, which can be modelled at all temperatures using a simple exponential relaxation function,

\begin{equation}
A(t)=A_{\mathsf{r}}\exp({-\lambda t})+A_{\mathsf{bg}}
\end{equation}

\noindent where ${\lambda}$ is the muon spin relaxation rate, $A_{\mathsf{r}}$ is the relaxing amplitude and $A_{\mathsf{bg}}$ is the background contribution from muons stopping outside the sample. Such an exponential relaxation of the LF signal of a sample is characteristic of electronic magnetic moment fluctuations within a motional narrowing regime with the corresponding relaxation rate described by the Redfield equation \cite{Redfield}: 

\begin{equation}
\lambda={\frac{2{\gamma}_{\mu}^{2}{\Delta}^2{\tau}}{1+{\gamma}_{\mu}^{2}B_{\mathsf{LF}}^{2}{\tau}^2}}
\end{equation}
  
\noindent where ${\gamma}_{\mu}$ is the muon spin gyromagnetic ratio, $B_{\mathsf{LF}}$ is the applied longitudinal-field and ${\Delta}$ and $\tau$ are the internal electronic magnetic field distribution and its fluctuation time---characteristic to a particular sample---respectively. A common feature that emerges from the analysis of LF ${\mu}$SR data of many QSL candidates of diverse chemical and structural varieties is the presence of a plateau in the muon spin relaxation rate, ${\lambda}$, at low-temperatures \cite{Gardner1999,Mendels2007,Clark2013,Li2016,Mustonen2018,Kenney2019}. Qualitatively, this is taken to indicate the onset of a QSL ground state characterised by dynamic electronic magnetic moment fluctuations with a temperature-independent fluctuation time. Quantitatively, however, it is challenging to meaningfully compare muon spin relaxation rates observed in the plateau states across different classes of QSL candidate materials because, as Equation 7 highlights, there are several key parameters that will determine the overall magnitude of ${\lambda}$ for any given system, as well as experimental factors, such as the applied LF and signal background. Even across a single family of materials---here Zn$_x$Cu$_{4-x}$(OH)$_6$FBr---substituting Cu$^{2+}$ for diamagnetic Zn$^{2+}$ will affect the size and distribution of the internal electronic magnetic field, which can have pronounced consequences for the magnitude of ${\lambda}$. However, for members of the Zn-barlowite series with $x>0.5$---for which we have demonstrated that magnetic order is suppressed---one would not expect the internal field distribution, ${\Delta}$, within a particular sample to vary significantly with temperature. Therefore, by normalising the muon spin relaxation rates obtained for highly-substituted members of the Zn-barlowite series, shown in Fig. S6 of the Supplementary Information, we can directly compare their dynamical magnetic moment correlations across the series, which will be primarily governed by their temperature-dependent field fluctuation time, $\tau$.\\

Fig. \ref{fig:5}(b) shows the comparison of the normalised relaxation rate, $\lambda_{\mathsf{norm}}$, for Zn$_x$Cu$_{4-x}$(OH)$_6$FBr with $x=0.66(1)$ and $0.99(1)$. Both samples have a prominent plateau in the muon spin relaxation that persists to the lowest measured temperatures---in the case of the $x=0.99(1)$ sample, to at least $50$~mK---indicating that their magnetic ground states are likely characterised by fluctuating Cu$^{2+}$ magnetic moments despite the strong exchange interactions between them ($|\theta|~\approx~190$~K for $x=0.99(1)$, see Fig. \ref{fig:2}(b)). Our observations are thus in support of the recent proposals of a QSL ground state for fully-subsituted ZnCu$_3$(OH)$_6$FBr from both experiment and theory \cite{Wei2019,Smaha2020}. Another important result from our present study, however, is the striking similarity of the temperature dependence of ${\lambda_{\mathsf{norm}}}$ for $x=0.66(1)$ and $0.99(1)$ samples, as seen in Fig. \ref{fig:5}(b). This suggests that the nature of the dynamical field fluctuations in highly-substituted Zn-barlowite---and its potential QSL state---is remarkably robust to the presence of interlayer exchange between the kagom{\'e} layers of its hexagonal crystal structure due to partial substitution at the interlayer sites or that the combination of chemically randomised magnetic interactions and geometric frustration results in a theoretically proposed disorder-induced QSL \cite{Kawamura2014a}. In either case, we conclude that $x~\approx~0.5$ marks a critical percolation threshold in the Zn-barlowite series, above which dynamic magnetic moment correlations characterise the low-temperature magnetic behaviour. We note that this is in agreement with a recent study which proposes that magnetic moments in single crystals of Zn-barlowite with $x=0.56$ remain dynamic to low-temperatures via bulk magnetometry measurements \cite{Smaha2020}.  

\section{Conclusions}

\noindent In summary, we have demonstrated that the Zn-barlowite series offers a promising route to a new QSL phase in an $S=\frac{1}{2}$ kagom{\'e} antiferromagnet. By combining ${\mu}$SR experiment with DFT, we find that the electronic magnetic moments in the parent material of this series, barlowite, \ch{Cu4(OH)6FBr}, are static below $T_{\mathsf{N}}=15$~K and that ${\mu}$SR spectra can be quantitatively interpreted by the formation of both $\mu$--F and $\mu$--OH complexes. We note that the application of DFT methods may be especially important in the $\mu$SR study of QSL candidates, as they allow us to assess not only the location of implanted muons but also the degree of distortion that they cause, thus providing confidence that the muon spin probes the intrinsic properties of the quantum disordered ground state \cite{Lancaster2018}. We find that incorporating as little as $x=0.16$ of Zn$^{2+}$ into the crystal structure of barlowite leads to a suppression of its ordered, frozen fraction, while for $x>0.66$ the electronic magnetic moments remain dynamically fluctuating at all measurable temperatures. Whether the magnetic ground states of such partially-substituted compositions of Zn-barlowite correspond to a QSL state---as proposed for the $x=1.0$ end-member \cite{Feng2017,Feng,Wei2019,Smaha2020}---awaits  further investigation, for instance, by inelastic neutron scattering. \\

Also of note is that we find no evidence in our $\mu$SR measurements for the coexistence of long-range magnetic order and dynamic magnetic moment fluctuations in barlowite, which has been previously observed by other groups via NMR measurements \cite{Ranjith2018} and in the structurally related clinoatacamite via $\mu$SR measurements \cite{Zheng2005}. Indeed, the reduced ordered moments obtained from the magnetic structure refinement of barlowite also indicate the presence of persistent dynamics in its ordered state \cite{Tustain2018}. However, an additional consideration here is the potential sample dependence on the magnetic properties of barlowite. Whilst the subtle structural differences depending on synthetic route are unlikely to have any significant effect on the muon stopping sites, some studies report several magnetic transitions in their samples synthesised by alternative routes to us which may eventually be insightful to explore using $\mu$SR \cite{Han2014,Pasco2018,Smaha2018,Smaha2020}.\\

To conclude this present study, we finish by noting the striking similarity of the temperature dependence of the normalised muon spin relaxation rates, ${\lambda}_{\mathsf{norm}}$, not only between members of the Zn-barlowite series, but also with members of the related Zn-paracamite series, and in particular, the QSL candidate herbertsmithite \cite{Mendels2007}, as shown in Fig. \ref{fig:5}(b). The apparent equivalence of the dynamics of the magnetic moment fluctuations that drive both of these related families of $S=\frac{1}{2}$ kagom{\'e} antiferromagnets into the plateau state is perhaps indicative of a universality to the way in which such hydroxyl halides evolve towards a QSL regime upon chemical substitution. Further direct comparison of the evolution of magnetic moment correlations in the Zn$_x$Cu$_{4-x}$(OH)$_6$FBr and Zn$_x$Cu$_{4-x}$(OH)$_6$Cl$_2$ series across complementary time and length scales by, for example, NMR spectroscopy or neutron spin-echo techniques, could, therefore, provide much-needed further experimental insight into the roles of interlayer coupling, cation occupancy disorder and local structure in the ground state selection of the $S={\frac{1}{2}}$ kagom{\'e} antiferromagnet.      

\section{Methods}

\noindent Polycrystalline samples of barlowite, \ch{Cu4(OH)6FBr}, were synthesised via a hydrothermal reaction as previously reported \cite{Tustain2018}. Zn-substituted samples, Zn$_x$Cu$_{4-x}$(OH)$_6$FBr, were achieved through the addition of excess \ch{ZnBr2} whilst varying the amount of \ch{CuBr2} accordingly as detailed in Table S1 of the Supplementary Information. For our fully Zn-substituted sample, $x=0.99(1)$, the autoclave was heated at $1$~K/min to $483$~K, held for $24$ hours and cooled at a rate of $0.1$~K/min to room temperature. The resulting products, coloured turquoise to pale blue depending on the level of Zn-substitution, were filtered and washed several times with distilled water and dried in air.\\

Sample purity was confirmed via powder X-ray diffraction (PXRD) and the Zn content determined using Inductively Coupled Plasma Optical Emission Spectrometry (ICP-OES). Time-of-flight PND measurements were performed on selected samples on the General Materials (GEM) diffractometer at the ISIS Facility of the Rutherford Appleton Laboratory \cite{GEM}. Magnetic susceptibility data were measured on a Quantum Design magnetic properties measurement system (MPMS) with a superconducting quantum interference device (SQUID) magnetometer in an applied field of $1$~T between $2$ and $300$~K.\\

$\mu$SR data were collected on the MuSR spectrometer at the ISIS Neutron and Muon Source \cite{MuSR,MuSR2}. Samples ranging from $60$ to $500$~mg and $x=0.00$ to $0.99(1)$ were packed into silver foil sachets, attached to a silver backing plate with vacuum grease and loaded into a $^{4}$He cryostat. For $x = 0.66(1)$ and $0.99(1)$, samples were additionally measured in a  $^{3}$He/$^{4}$He dilution fridge. Data were collected in zero-field (ZF), longitudinal-field (LF) and transverse-field (TF) geometries and analysed using the MANTID software \cite{Arnold2014}. A separate sample of \ch{Cu4(OH)6FBr} was measured on the General Purpose Surface-Muon (GPS) instrument of the Swiss Muon Source at Paul Scherrer Institut. The 90 mg sample was contained in an aluminium foil packet, suspended in the muon beam using a silver fork-type holder and measured in VETO mode to reduce background signal. DFT muon-site calculations were carried out using the MuFinder software \cite{Huddart2020} and the plane-wave-based code CASTEP \cite{Clark2005} using the local density approximation. A supercell consisting of $1 \times 1 \times 2$ unit cells of the $Pnma$ structural model of barlowite was used in order to minimise the effects of muon self-interaction resulting from the periodic boundary conditions. Muons, modelled by an ultrasoft hydrogen pseudopotential, were initialised in low-symmetry positions and the structure was allowed to relax (keeping the unit cell fixed) until the change in energy per ion was less than $1 \times10^{-5}$ eV. A cutoff energy of $544$ eV and a $1 \times 2 \times 1$ Monkhorst-Pack grid \cite{Monkhorst1976} were selected for $k$-point sampling.

\section{Data Availability}

\noindent Data supporting the findings within this study are available from the corresponding author upon request. Raw data sets from ISIS experiments can be accessed via links provided in Refs. \cite{GEM,MuSR,MuSR2}. \\

\section{Acknowledgements}

\noindent L.~C. and K.~T. acknowledge the University of Liverpool for financial support and a studentship to K.~T. B.~M.~H. thanks the STFC for the provision of a studentship. T.~-H.~H. acknowledges the support of the Grainger Fellowship provided by the University of Chicago. Work at the ISIS Neutron and Muon Source was supported by the STFC. Part of this work is supported by the EPSRC grant no EP/N024028/1. F.~B. acknowledges support from the French Agence Nationale de la Recherche under Grant No. ANR-18-CE30-0022 LINK and from the Universit{\'e} Paris-Sud (MRM MAGMAG grant). The authors are grateful to S.~Moss for performing ICP-OES measurements, H.~Niu and B.~Le~Pennec for assistance with susceptibility measurements, as well as I.~da~Silva for assistance on GEM. We acknowledge computing resources provided by Durham Hamilton HPC.}

\section{Author Contributions}

\noindent K.~T., B.~W.~-O'~B. and T.~-H.~H. prepared the samples. K.~T., F.~B., P.~J.~B. and L.~C. performed the MuSR experiments and K.~T. analysed the data. F.~B. and H.~L. performed the GPS experiments and F.~B. analysed the data. T.~L. and B.~M.~H. performed the DFT calculations. K.~T. performed supplementary experiments (magnetic susceptibility and PND) and analysed the data. L.~C. conceived of and supervised the study. L.~C. and K.~T. wrote the manuscript with input from all of the authors. The authors declare no competing interests.

\widetext
\clearpage
\begin{center}
\textbf{Supplementary Information for From magnetic order to quantum disorder: a $\mu$SR study of the Zn-barlowite series of $S={\frac{1}{2}}$ kagom{\'e} antiferromagnets, Zn$_{x}$Cu$_{4-x}$(OH)$_{6}$FBr}
\end{center}

\setcounter{figure}{0}
\setcounter{table}{0}
\renewcommand{\thefigure}{S\arabic{figure}}
\renewcommand{\thetable}{S\arabic{table}}

\begin{figure}[h]
\centering
\includegraphics[width=\linewidth]{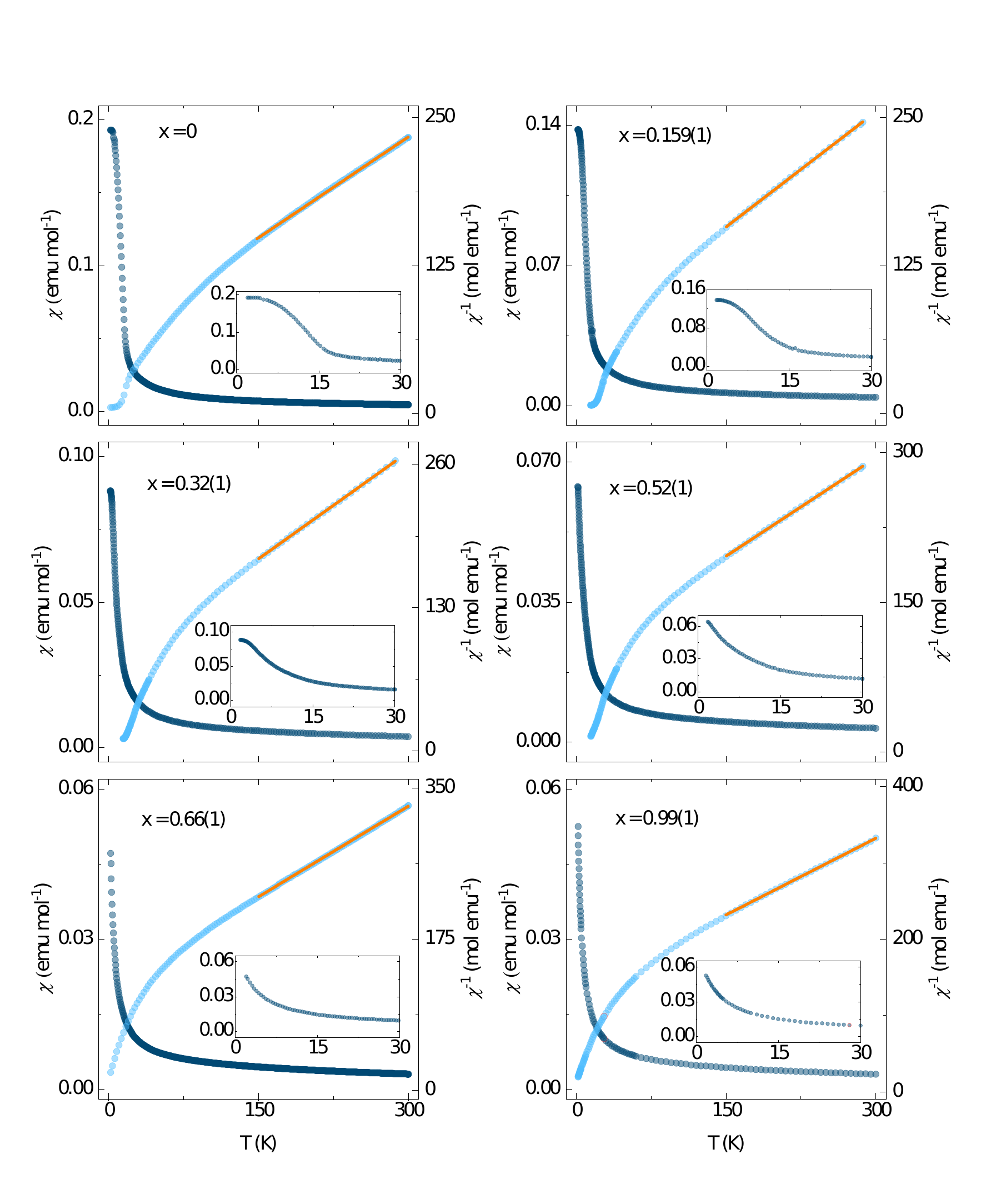}
\caption{The temperature dependence of the magnetic susceptibility for samples of Zn-barlowite, \ch{Zn_{\textit{x}}Cu_{4-\textit{x}}(OH)6FBr}. Fits to the inverse susceptibility using the Curie-Weiss law are shown in orange, and the resulting parameters for each fit are shown in Fig. 2(b) of the main manuscript.}
\label{fig:S1}
\end{figure}

\begin{figure}[h]
\centering
\includegraphics[width=\linewidth]{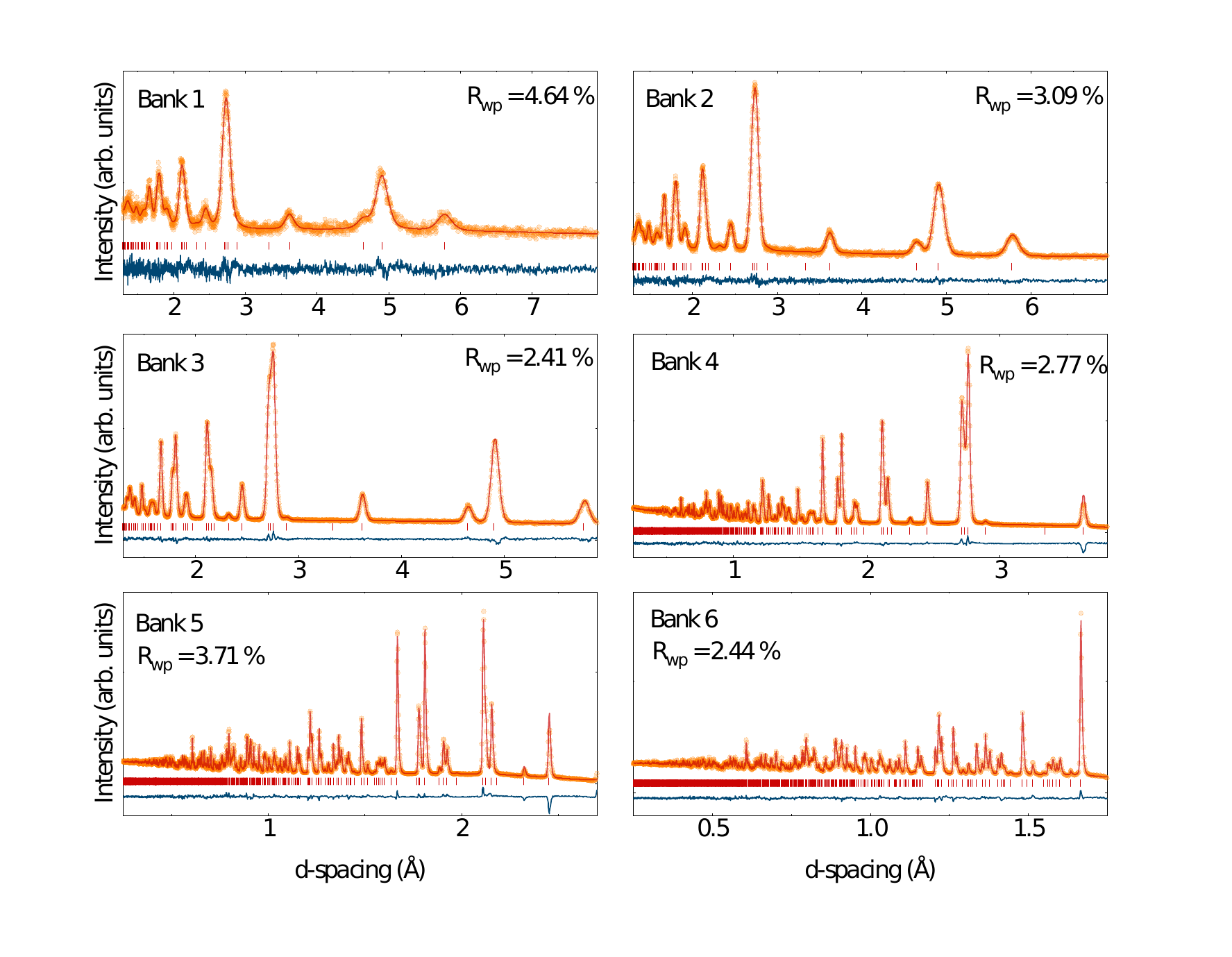}
\caption{Powder neutron diffraction data collected at 2~K on Bank 4 of GEM and Rietveld refinement of the $P6_{3}/mmc$ model for \ch{Zn_{0.60(1)}Cu_{3.40}(OD)_6FBr}. Data points are shown in orange, the fitted curve is shown in red and Bragg peak positions are represented by dark red tick marks. The refined fractional occupancies for the intralayer (Cu1/Zn1) and interlayer (Cu2/Zn2) sites are 0.996(3)/0.004 and 0.412/0.588(9), respectively. $R_{wp}$ = 2.41~\% and $\chi^{2}=2.74$.
}
\label{fig:S2}
\end{figure}

\begin{figure}[h]
\centering
\includegraphics[width=\linewidth]{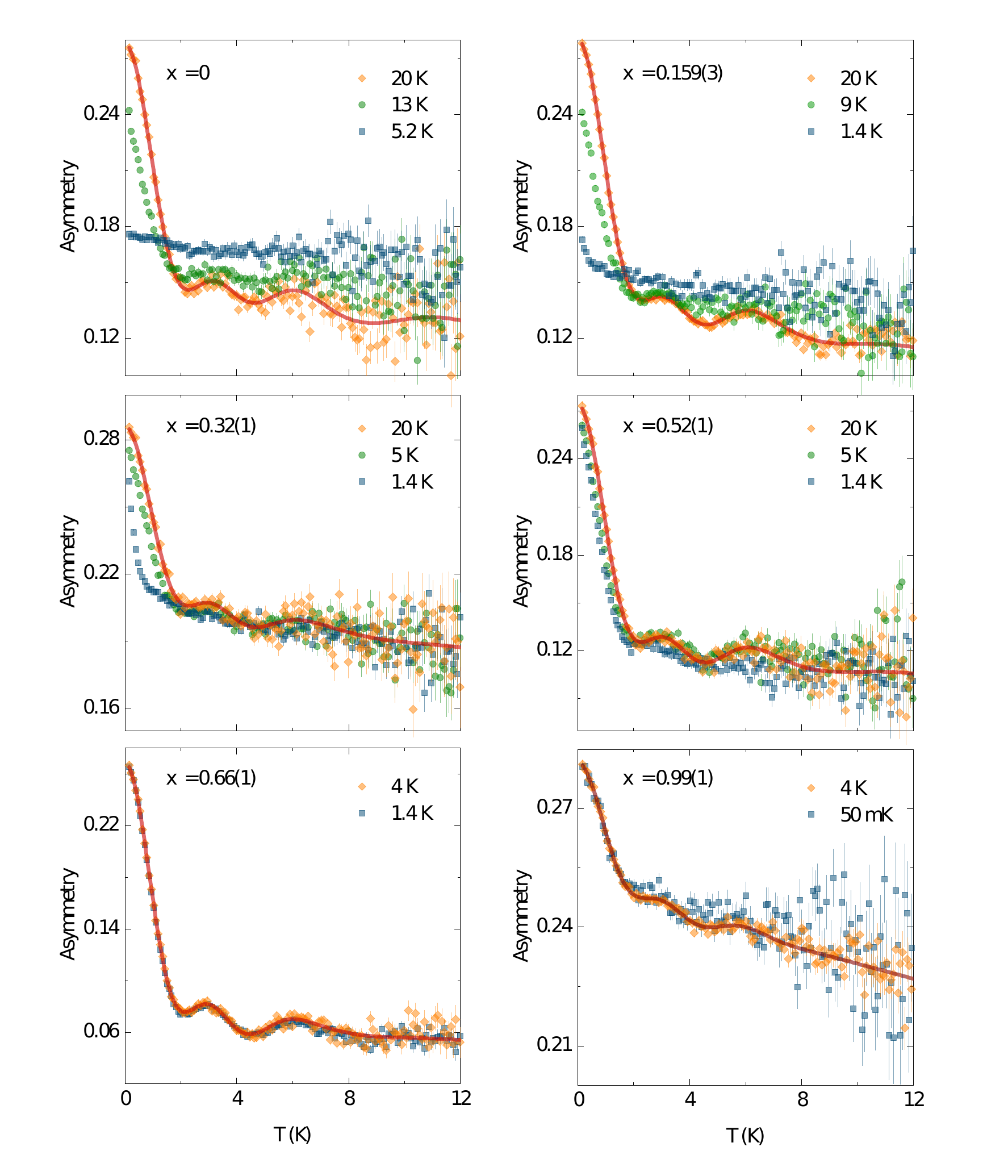}
\caption{Asymmetries measured in zero field for Zn-barlowite, \ch{Zn_{\textit{x}}Cu_{4-\textit{x}}(OH)6FBr}. The red lines indicate fits of Equation 1 in the main text, and parameters obtained are shown in Table \ref{table:S3}.}
\label{fig:S3}
\end{figure}

\begin{figure}[h]
\centering
\includegraphics[width=\linewidth]{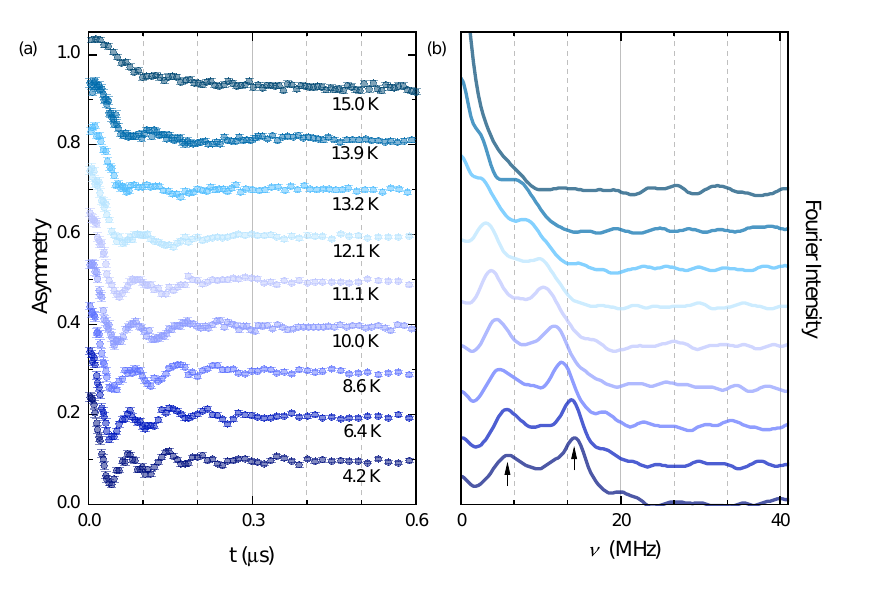}
\caption{(a) Asymmetries collected on the GPS instrument at short times in zero-field and at various temperatures for \ch{Cu4(OH)6FBr} below $T_{\mathsf{N}}=15$~K. (b) Fourier transform of the asymmetries shown in the left panel. The black arrows indicate the two main frequencies at 4.2~K. Curves are shifted offset for clarity.}
\label{fig:S4}
\end{figure}

\begin{figure}[!h]
\centering
\includegraphics[width=0.6\linewidth]{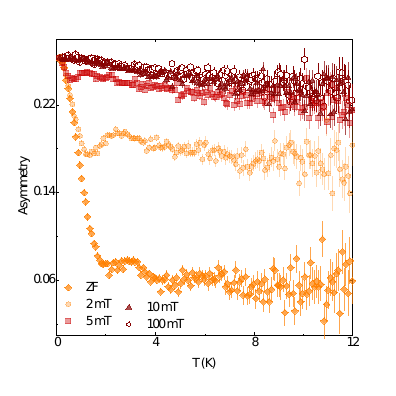}
\caption{ The field dependence of the muon spin depolarisation at 0.15~K for \ch{Zn_{0.66(1)}Cu_{3.34}(OH)6FBr}.}
\label{fig:S5}
\end{figure}

\begin{figure}[!b]
\centering
\includegraphics[width=\linewidth]{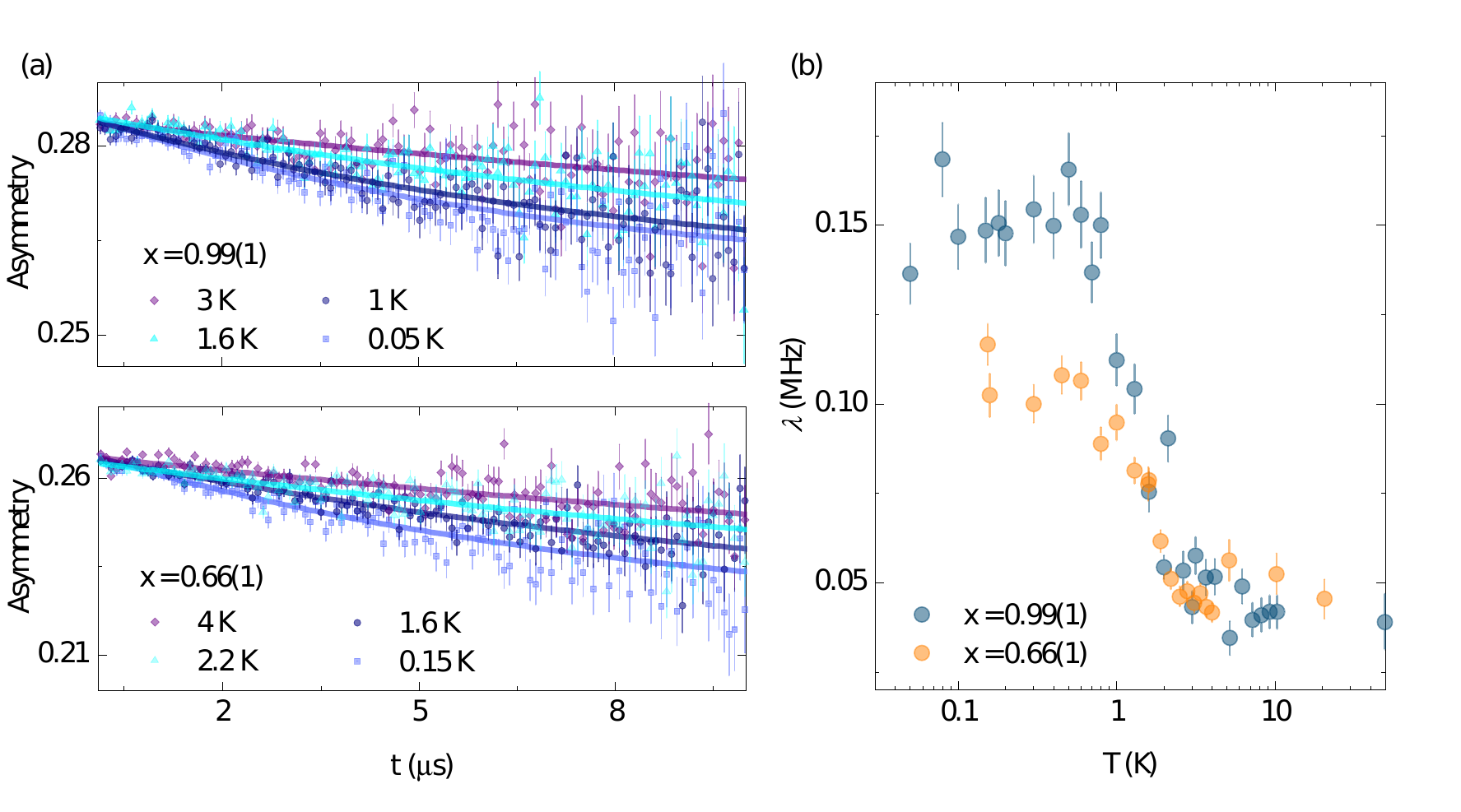}
\caption{(a) Time-dependent muon asymmetry measured in a 10~mT longitudinal-field for \ch{Zn_{0.99(1)}Cu_{3.01}(OH)6FBr} (top) and \ch{Zn_{0.66(1)}Cu_{3.34}(OH)6FBr} (bottom). Solid lines are fits using a simple exponential. (b) The resulting temperature dependence of the relaxation rate, $\lambda$.}
\label{fig:S6}
\end{figure}

\noindent Table \ref{table:S1} shows the quantities of reactants used to synthesise the Zn$_{x}$Cu$_{4-x}$(OH)$_{6}$FBr series. Fig. \ref{fig:S1} shows the temperature dependence of the magnetic susceptibility along with Curie-Weiss fits to the inverse susceptibility for all samples included in the main manuscript. Parameters obtained from these fits are shown in Fig. 2(b) of the main text. For $x=0.66(1)$ and $=0.99(1)$, we observe a Curie tail at low temperatures, which may be due to paramagnetic \ch{Cu^{2+}} defects. A deuterated sample of Zn-barlowite was synthesised using 6~mmol \ch{CuCO3.Cu(OH)2}, 2~mmol \ch{CuBr2}, 6~mmol \ch{ZnBr2}, 5~mmol \ch{NH4F} and 20~mL \ch{D2O} and the reaction mixture was heated in a 50-mL autoclave as described in Table \ref{table:S1}. Fig. \ref{fig:S2} shows powder neutron diffraction data collected for the resulting sample, \ch{Zn_{0.60(1)}Cu_{3.40}(OD)_6FBr}, at 2~K. We find that the crystal structure of this sample of Zn-barlowite is best described by the hexagonal $P6_{3}/mmc$ model at all temperatures, which also describes the room temperature $x=0$ parent material [23, main manuscript].
\\
\\
\noindent
Fig. \ref{fig:S3} shows the ZF muon asymmetry measured on MuSR for Zn$_{x}$Cu$_{4-x}$(OH)$_{6}$FBr above and below $T_{\mathsf{N}}=15$~K of the $x=0$ parent compound. Solid lines show fits of Equation 1 in the main text to  high-temperature data. The resulting parameters, shown in Table \ref{table:S3}, show little variation across the series indicating no change in the muon stopping sites. As \textit{x} increases, the loss of initial asymmetry also increases. For $x=0.52(1)$ the frozen fraction at low temperature, shown in Fig. 4(c) of the main text, has been estimated by taking the asymmetry at $t=0.1$~$\mu$s as a proportion of the full asymmetry in the paramagnetic regime. For $x=0.66(1)$ and $x=0.99(1)$, the rate of depolarisation at base temperature is comparable to the rate of depolarisation at 4~K and the asymmetry data relax to the same baseline at all temperatures. Note that for $x=0.99(1)$ the relaxing tail at long times is a result of a high Ag background during this measurement due to the small sample size. Fig. \ref{fig:S4} shows the ZF muon asymmetry measured on the GPS instrument and the corresponding Fourier transforms for \ch{Cu4(OH)6FBr} below $T_{\mathsf{N}}=15$~K. Fig. \ref{fig:S5} shows the LF field dependence of the muon asymmetry for $x=0.66(1)$ at 150~mK, where it is evident that a field of 10~mT is sufficient to decouple the muon spin from the nuclear magnetic moments. Asymmetries measured for $x=0.66(1)$ in an applied longitudinal field of 10~mT are shown in Fig. \ref{fig:S6} along with the temperature dependence of $\lambda$ obtained from fits of Equation 6 in the main manuscript. The normalised relaxation rates, $\lambda_{\mathsf{norm}}$, shown in Fig. 5 of the main manuscript, were obtained by normalising the data shown in Fig. \ref{fig:S6}(b) between 0 and 1 with respect to the maximum and minimum values of $\lambda$ within each data set. \\ \\

\begin{table}[h]
\caption{Summary of the reaction quantities used for the synthesis of Zn$_{x}$Cu$_{4-x}$(OH)$_{6}$FBr. Note that all reactants were heated at 10~K/min to 473~K and held for 72~h and cooled to room temperature at 5~K/min, except for $x=0.99(1)$ where the heating profile is detailed in the main manuscript.}
\begin{ruledtabular}
\hspace{-0.75cm}
\begin{tabular}{@{}ccccccc}
$x$ & \ch{CuCO3.Cu(OH)2} / mmol & \ch{CuBr2} / mmol & \ch{ZnBr2} / mmol & \ch{NH4F} / mmol & \ch{H2O} / mL & Autoclave / mL\\
\hline \vspace{-0.2cm} \\
0 & 4 & 8 & 0 & 12 & 20 & 50\\
0.159(3) & 2 & 2 & 2 & 2 & 20 & 50\\
0.32(1) & 1 & 1.5 & 0.5 & 4 & 10 & 23\\
0.52(1) & 2 & 2 & 4 & 4 & 20 & 50 \\
0.66(1) & 2 & 1 & 3 & 4 & 10 & 23\\
0.99(1) & 0.5 & 0.25 & 0.75 & 1 & 2 & 23\\
\vspace{-0.3cm}
\label{table:S1}
\end{tabular}
\end{ruledtabular}
\end{table}

\begin{table}[h]
\caption{\label{label}Rietveld refinement structure parameters for the $P6_{3}/mmc$ model fitted to powder neutron diffraction data collected at 2~K for Zn$_{0.60}$Cu$_{3.40}$(OD)$_{6}$FBr. Refined lattice parameters are $a=b=6.6638(5)$~\AA{} and $c=9.2861(7)$~\AA{}.}  
\begin{ruledtabular}
\hspace{-0.75cm}
\begin{tabular}{@{}lclllll}

Atom & Site & $x$ & $y$ & $z$ & Occupancy & $U_{\mathsf{iso}}$ (\AA$^{2}$)\\
\hline \vspace{-0.2cm} \\
Cu1 & $4a$ & 0.5 & 0 & 0 & 0.996(3) & 0.00326(7)\\
Zn1 & $4a$ & 0.5 & 0 & 0 & 0.004(3) & 0.00326(7)\\
Cu2 & $2d$ & 0.3333 & 0.6667 & 0.75 & 0.412(9) & 0.00278(4)\\
Zn2 & $2d$ & 0.3333 & 0.6667 & 0.75 & 0.588(9) & 0.00278(4)\\
F & $4c$ & 0 & 0 & 0.75 & 1 & 0.0094(2)\\
Br & $4c$ & 0.6667 & 0.3333 & 0.75 & 1 & 0.0027(1)\\
O1 & $8d$ & 0.20217(4) & 0.79783(4) & 0.90756(4) & 1 & 0.00416(7)\\
D1 & $8d$ & 0.12442(4) & 0.87558(4) & 0.86572(4) & 1 & 0.0147(1)
\label{table:S2}

\end{tabular}

\end{ruledtabular}
\end{table}

\begin{table}[h]
\caption{\label{label}Parameters obtained from a fit of Equation 1 to data collected in zero-field for \ch{Zn_{x}Cu_{4-x}(OH)6FBr}, shown in  Fig. \ref{fig:S3}. $f$ describes the fraction of muons at each of the two stopping sites, $\mu$--F and $\mu$--OH. $\omega$ is the angular frequency used to calculate the $\mu$-F or $\mu$-OH distances, $d$, and $\Delta$ describes the distribution of fields at each site induced by dipolar coupling to nearby nuclei.}

\begin{ruledtabular}
\hspace{-0.75cm}
\begin{tabular}{@{}ccc|ccccc|cccc}
~&~&~&\multicolumn{4}{c}{$\mu$--F}&~&\multicolumn{4}{c}{$\mu$--OH}\\
$x$&$T$ (K)&&~ $f$ (\%) & $\omega$ (Mrad s$^{-1}$) & $\Delta$ (mT) & $d$ &~& $f$ (\%) & $\omega$ (Mrad s$^{-1}$) & $\Delta$ (mT) & $d$ (\AA{}) 
\vspace{0.2cm}
\\
\hline \vspace{-0.2cm} \\
0&20&~& 69(4) & 1.23(2) & 0.43(3) & 1.22(1)&~ & 31(4) & 0.72(2) & 0.18(3)& 1.50(3)\\
0.159(3)&20&~& 65(2) & 1.25(1) & 0.44(1) & 1.22(1)&~ & 35(1) & 0.68(1) & 0.22(1)& 1.52(1)\\
0.32(1)&20&~& 70(8) & 1.24(4) & 0.48(1) & 1.22(3)&~ & 30(6) & 0.63(1) & 0.25(9)& 1.56(9)\\
0.52(1)&20&~& 66(4) & 1.25(2) & 0.45(3) & 1.22(1)&~ & 34(2) & 0.67(2) & 0.23(4)& 1.53(4)\\
0.66(1)&4&~& 70(2) & 1.30(1) & 0.48(2) & 1.20(1)&~ & 30(2) & 0.67(1) & 0.27(2)& 1.54(2)\\
~&1.4&~& 72(2) & 1.29(1) & 0.50(1) & 1.20(1)&~ & 28(1) & 0.65(1) & 0.30(2)& 1.53(2)\\
0.99(1)&4&~& 61(7) & 1.32(3) & 0.41(5) & 1.20(2)&~ & 39(6) & 0.69(4) & 0.31(8)& 1.51(7)\\
\vspace{-0.6cm} \\

\label{table:S3}
\end{tabular}

\end{ruledtabular}
\end{table}

\end{document}